\documentclass[12pt]{article}
\usepackage{amsmath, setspace}
\usepackage{graphicx}
\usepackage{lineno}

 \makeatletter       
 \renewcommand{\@biblabel}[1]{#1.}
 \makeatother
\begin{document}
\noindent
\begin{center}
{\LARGE\textbf{--Preliminary Report--}}
\end{center}
{\LARGE\textbf{Enceladus Farm: Can plants grow with Enceladus' water?}}
\\
\\
\begin{center}
{\large Daigo Shoji$^1$}
\end{center}
1. Earth-Life Science Institute, Tokyo Institute of Technology, 2-12-1 Ookayama, Meguro-ku, Tokyo (shoji@elsi.jp)\\
\begin{center}
\section*{Abstract}
\end{center}
Enceladus is a saturnian satellite that should have liquid water inside of it (subsurface ocean). Measurements and experiments on water plume from Enceladus have revealed that Enceladus' ocean contains several salts such as NaCl. On the Earth, salt in soil has become a serious problem for agriculture, and importance of salt-tolerant plants are indicated. In order to test the effect of Enceladus' water to terrestrial plant, by hydroponic, we tried to grow three salt-tolerant plants (ice plant, swiss chard and salicornia) simulating Enceladus' water (Enceladus Farm project). Using water with 0.33\% NaCl and 0.4\% NaHCO$_3$, which is consistent with the observations of Enceladus, all plants could grow if they were germinated and grown with pure water until each plant had a few leaves. However, growth rate can be suppressed compared with the plants cultivated with pure water. Because our first test was performed with loose conditions, more works are needed to evaluate the effect of Enceladus' water to plant growth. However, in addition to the works to grow plants on lunar and martian grounds, Enceladus' water may be used to consider properties of plant from wider environment.
\\
\\
\section{Introduction}
It has been observed that Enceladus, one of the mid-sized saturnian satellites, emits water plumes to the outer space (Porco et al., 2006). In the plume, salts and nano-silica have also been detected, which strongly implies the interaction between liquid water and silicate core (Postberg et al., 2009, 2011; Hsu et al., 2015) Although the detail is still in debate, due to these observations, Enceladus is one of the most probable bodies that have a subsurface ocean. Liquid water is a key factor for generation and habitability of organisms. Thus, Enceladus is also an important object for astrobiology as well as planetary science.

In the field of space engineering, it has been tested whether plants can grow in soil of Moon and Mars (Wamelink et al., 2014). When people stay on such bodies in future, one serious problem is how we produce foods. If we can grow plants using lunar and martian soil directly, it can be one solution. Unfortunately, although Enceladus is an attractive body for astrobiology, contrast to Moon and Mars, it has not been tested whether plants can grow with Enceladus' water. This is not surprising because, compared with Moon and Mars, Enceladus is not a practical target for humans to live for a long time.

However, liquid water in Enceladus may be used to analyze the properties of plants on the Earth. Observations found that Enceladus' ocean contains several salts such as NaCl (Postberg et al., 2009, 2011). Recently, salinization in terrestrial ground has been a very serious problem for agriculture and ecosystem (e.g., Rozema and Flowers, 2008). Typical plants cannot live when salt is mixed in soil. Thus, when salinity increases due to irrigation or environment change, conventional agriculture cannot be maintained there. 
One solution to the salinized soil is to grow salt-tolerant plants (Rozema and Flowers, 2008). As mentioned above, typical plants cannot live in high salinity soil. However, several plants have the mechanism to grow in the salty environment. If we can grow salt-tolerant plants, agriculture may be possible even on salinized region. In order to test the salt tolerance of plants, many researches have been performed (e.g., Lv et al., 2012). In these works, water and soil conditions are tested focusing on terrestrial environment. Considering terrestrial agriculture, it is not surprising. However, in space engineering, tests of plant growth have been performed considering the soil conditions of Moon and Mars. Rocky and icy bodies in our solar system have diverse environment. Thus, if we can compare plant growth between several bodies' environment, it may be valuable for the terrestrial agricultural problem as well as space engineering. 

Due to this motivation, we started a new project to grow plants simulating Enceladus' water, which is named "Enceladus Farm". As a first test, we grew three salt-tolerant plants, ice plant, salicornia and swiss chard using NaCl (table salt) and NaHCO$_3$ (baking soda). Because this test is very loose, more works are required in order to check the plant growth. However, in this work, preliminary results of our first test are shown.

\section{Method}
\subsection{Water composition}
\begin{figure}[htbp]
\centering
\includegraphics[width=13cm]{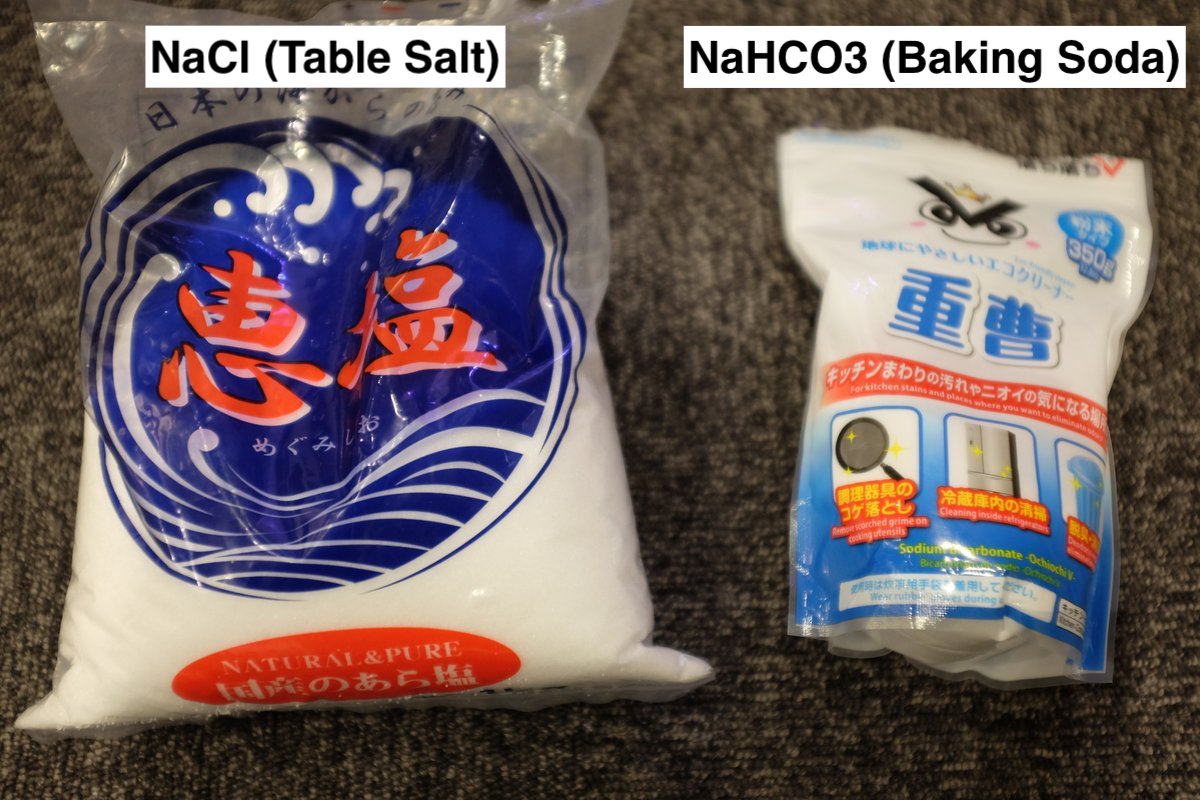}
\caption{For the simulation of Enceladus' ocean, as a first step, only NaCl and NaHCO$_3$ were used. Both materials were obtained from market.}
\label{salt}
\end{figure}

Measurements by Cassini probe revealed that Enceladus' plume is mainly composed of three types of particles (type I: almost pure water ice, type II: significant amount of organic and/or siliceous material, type III: ice with high sodium content) (Postberg et al., 2009, 2011). Of the three types, high sodium particle (type III) was formed by freezing of water droplet, which can be the information on composition of the subsurface ocean. Postberg et al. (2009) performed experiments on ion spectrum from salt water. By their experiments, the observed spectrum is consistent with the water with 0.05-0.2 mol kg$^{-1}$ NaCl and 2-5 times more NaCl than NaHCO$_3$ (Postberg et al., 2009). In order for nano-silica to be formed by cooling of hydrothermal water, pH of Enceladus' ocean should be 8.5-10.5, which implies that the maximum salinity is 4\% at the boundary between the ocean and the core (Hsu et al., 2015). Although other salts and chemical materials are contained in Enceladus, for simplicity, we simulated Enceladus' water only by NaCl and NaHCO$_3$, which were obtained from market (Fig. \ref{salt}). Salinities of NaCl and NaHCO$_3$ were set at 0.33 \% and 0.4 \%, respectively, which are consistent with Postberg et al. (2009). Both materials were dissolved into tap water. One caveat is that, in Enceladus' ocean, ammonia is also contained (Hsu et al., 2015). In this work, as the first test, we ignored ammonia. Thus, pH of the water was $\sim$8, which is lower than Enceladus' ocean (pH=8.5-10.5) (Hsu et al., 2015). The other caveat is that, in addition to NaCl and NaHCO$_3$, we added liquid fertilizer to the salt water because plants cannot grow without basic ingredient such as nitrogen. 

\subsection{Cultivation}
\begin{figure}[htbp]
\centering
\includegraphics[width=17cm]{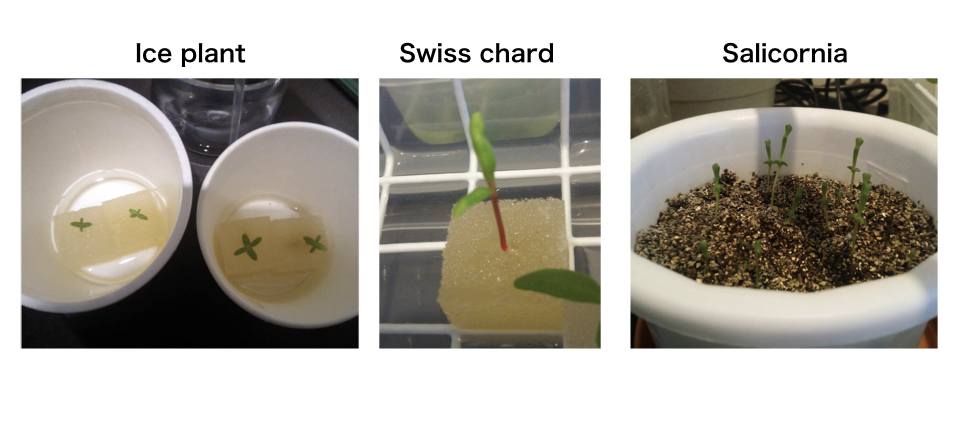}
\caption{Sprouts of three salt-tolerant plants germinated with pure water. For salicornia, vermiculite was used because it did not germinate well on sponge.}
\label{germination}
\end{figure}

Because the purpose of this first test is to check whether salt-tolerant plants can grow with Enceladus-like salt water, as the loose condition, germination of seeds was conducted with pure (tap) water without salts. If the germinated plants can grow, germination with salt water can be a future work. For the germinations of ice plant and swiss chard, seeds were put on cubic sponges (2.5 cm each side) that has cuts at their center (Fig. \ref{germination}). Seeds of salicornia were put to vermiculite because they did not germinate well on the sponge (Fig. \ref{germination}).

\begin{figure}[htbp]
\centering
\includegraphics[width=15cm]{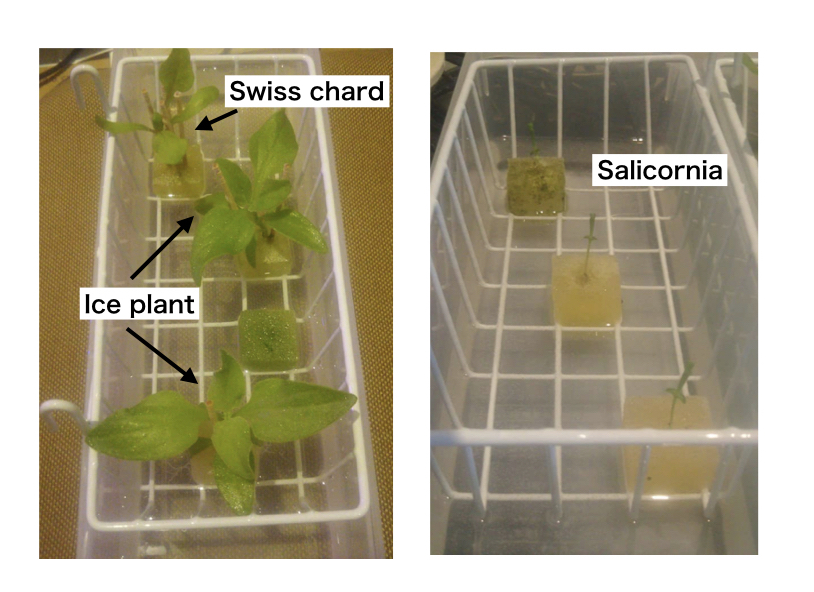}
\caption{Sprouts moved to salt water tank. When germinated ice plant and swiss chard had a few leaves and salicornia became a few centimeters in height, they were moved to the salt water tank. }
\label{salt_bath}
\end{figure}

\begin{figure}[htbp]
\centering
\includegraphics[width=15cm]{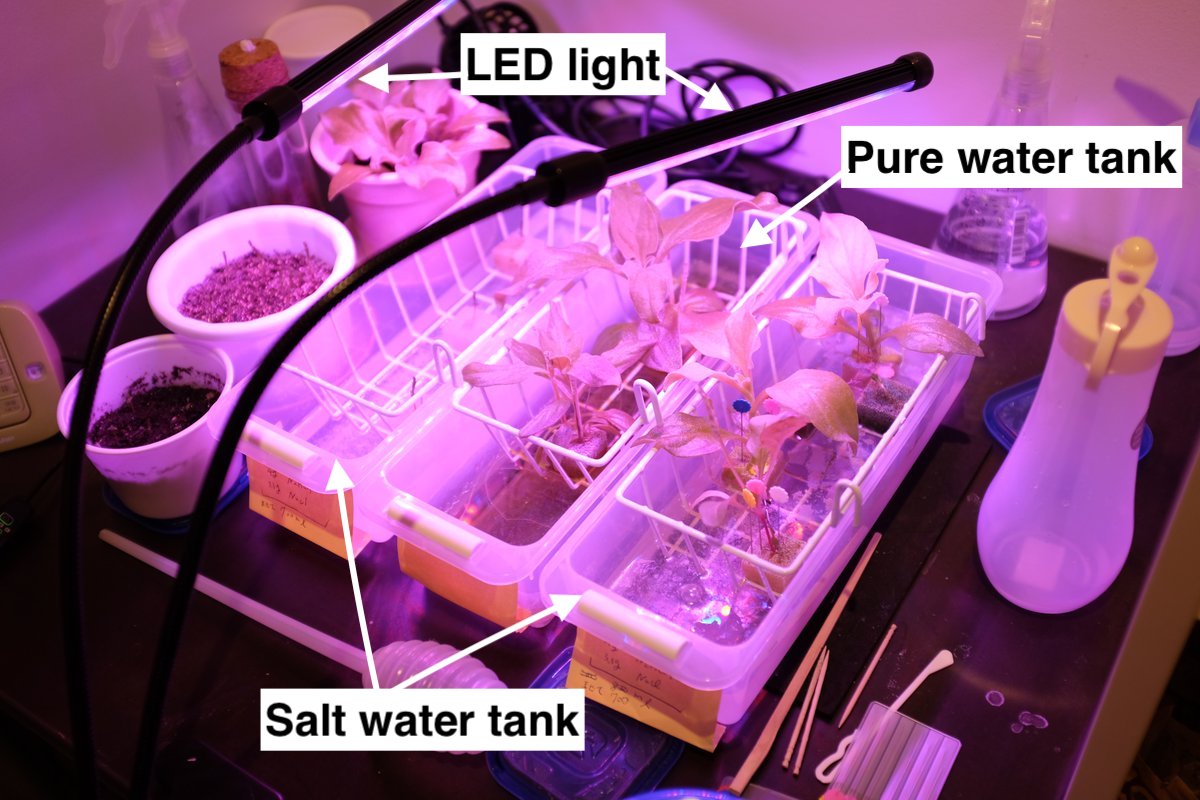}
\caption{Plants cultivated in salt water tanks. As a comparison, two ice plants were cultivated with pure water tank (tap water + liquid fertilizer). Every tank was set in room and LED lights were used for photosynthesis. }
\label{tank}
\end{figure}

When each plant has a few leaves (when height became 2-3 cm in the case of salicornia), they were moved to the water tank filled with salt (Enceladus-like) water (Fig. \ref{salt_bath}). Tanks were put in room, and LED lights were used for nine hours per day (Fig. \ref{tank}). The salt water was changed once a week. However, when the amount of water decreased within one week, pure water with liquid fertilizer was filled in the tanks.

\section{Results}
\subsection{Ice plant}
\begin{figure}[htbp]
\centering
\includegraphics[width=15cm]{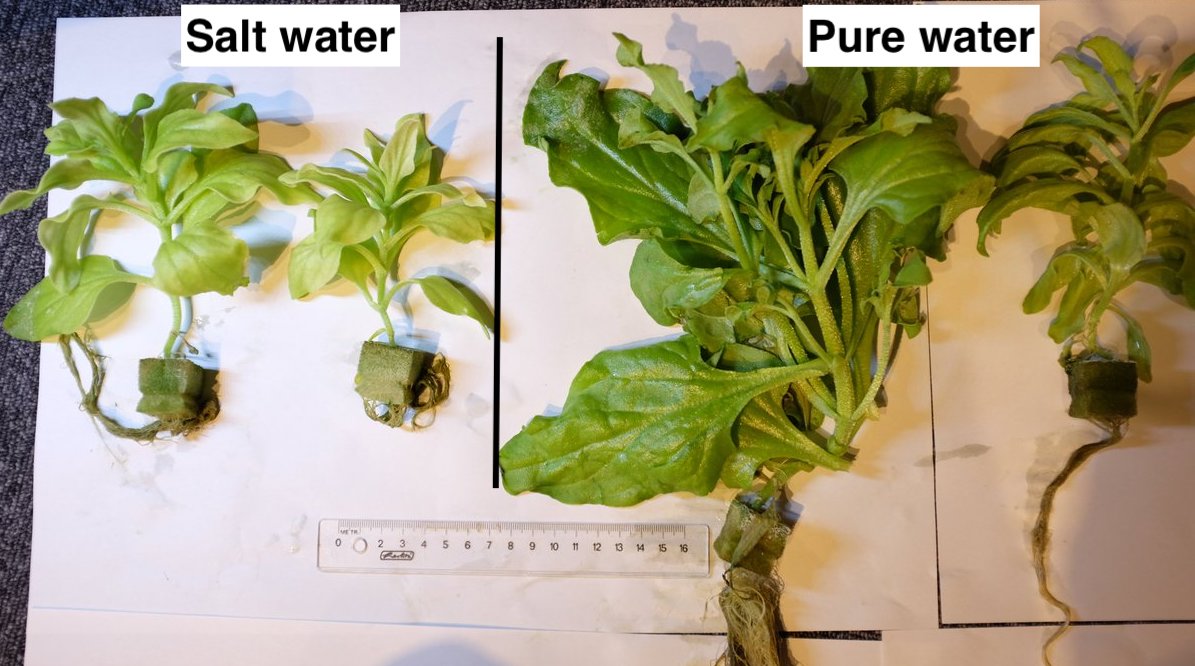}
\caption{Ice plants about one and a half months after they were moved to the salt water tank. Two plants on the right side are the ones cultivated in the pure water tank (One plant was moved to the salt water at Jan. 24, 2020, and then one more plant was moved at Jan. 31, 2020. The photo was taken at Mar. 19, 2020).}
\label{ice_harvest}
\end{figure}

We grew two sprouts of ice plant in the salt water tank. For a comparison, two sprouts were also cultivated only with pure(tap) water and liquid fertilizer. Fig. \ref{ice_harvest} shows the ice plants $\sim$1.5 months after the movement to the salt water tank. Both ice plants could grow and new leaves were generated with the salt water. However, compared with the plants grown in the pure water tank, growth rate was reduced. The height of ice plants from the salt water tank was $\sim$11 cm and $\sim$8 cm while the plants from the pure water tank had $\sim$17 cm and $\sim$13 cm, respectively. The color of leaves also became lighter at the salt water. Although ice plant is salt-tolerant species, chemical materials of Enceladus' water seem to suppress its growth.

\subsection{Swiss chard}
\begin{figure}[htbp]
\centering
\includegraphics[width=13cm]{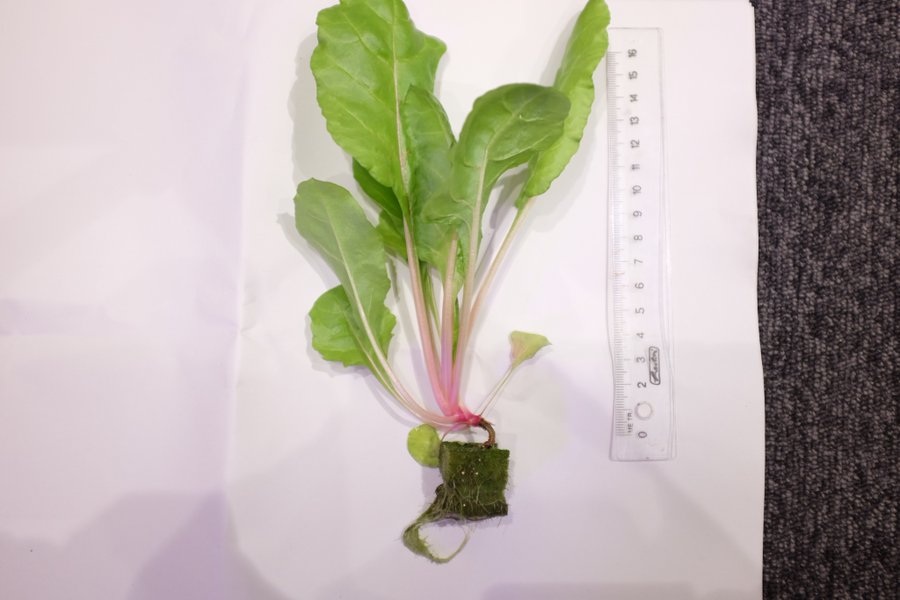}
\caption{Swiss chard about one and a half months after they were moved to the salt water tank (the plant was moved to the salt water at Jan. 31, 2020 and the photo was taken at Mar. 24, 2020). Because the base of the stem became brown, it was moved to pure water between Feb. 20, 2020 and Mar. 5, 2020.}
\label{chard_harvest}
\end{figure}

Swiss chard also grew in the salt water tank. However, contrast to the progressive growth of leaves, the base of the stem became brown. Although the reason of the color change is uncertain, when swiss chard grew upward, it fell several times. Thus, the stem might be damaged due to the falls and the salt in water. Due to this color change, we moved the plant to pure water, and then new roots were generated upside of the brown stem. Because the chard began to grow further with pure water, after two weeks, it was moved to the salt water tank again. Although it was cultivated with pure water for two weeks, for one and a half months, the height of the chard became from a few centimeters to $\sim$20 cm with the salt water (Fig. \ref{chard_harvest}). 

\subsection{Salicornia}
\begin{figure}[htbp]
\centering
\includegraphics[width=9cm]{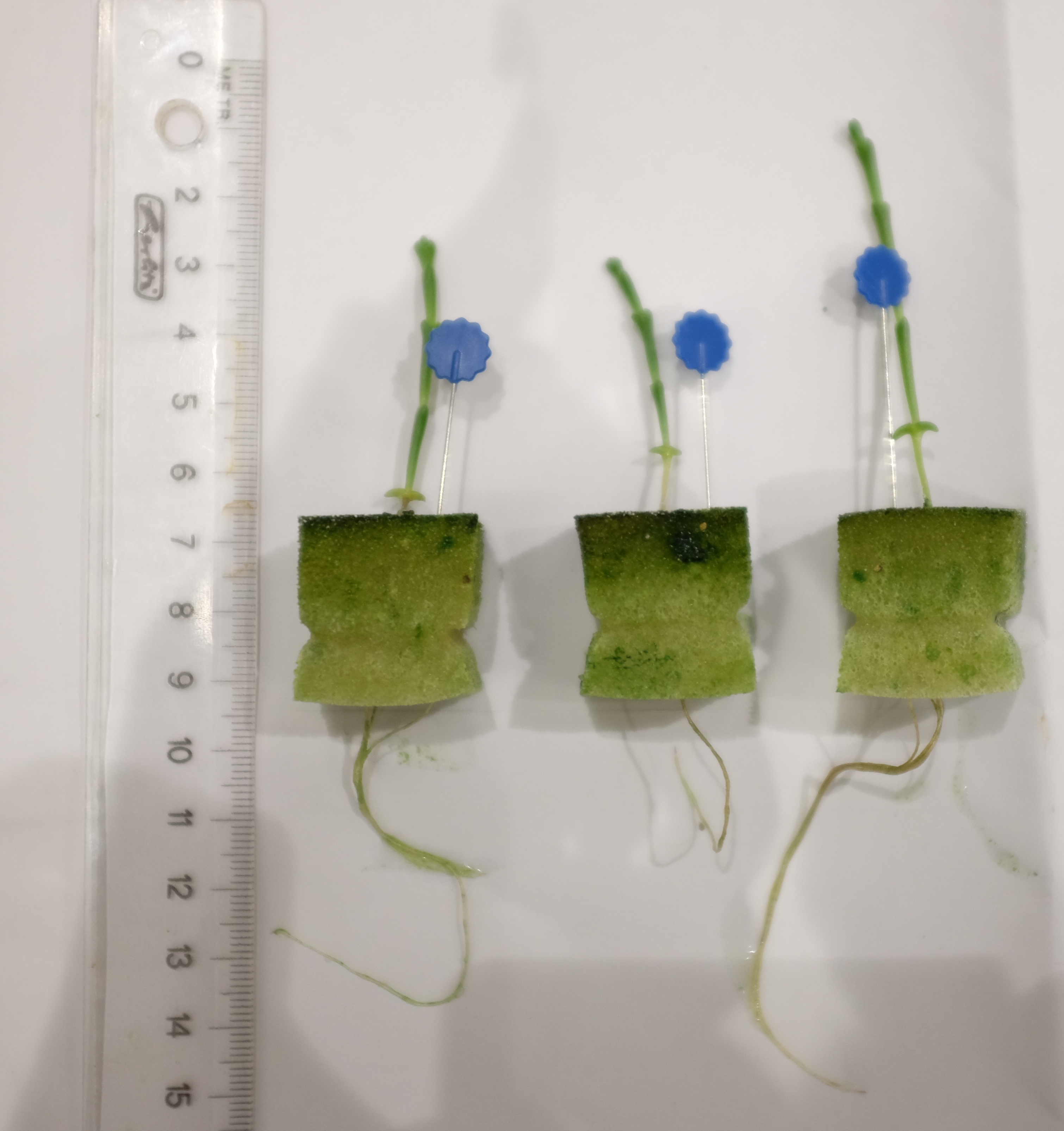}
\caption{Salicornia about one and a half months after they were moved to the salt water tank (the plants were moved to the salt water at Feb. 7, 2020 and the photo was taken at Mar. 26, 2020).}
\label{salicornia_harvest}
\end{figure}

We moved three germinated sprouts of salicornia from vermiculite to the salt water tank. Although salicornia also grew with the salt water, in our cultivation, the growth speed was slow. For $\sim$1.5 months, the height of salicornia became $\sim$5 cm (approximately 4.5 cm, 4.0 cm, 6 cm, respectively) (Fig. \ref{salicornia_harvest}). It has been indicated that salicornia is one of the strongest plants to salt (e.g., Lv et al., 2012). In addition to the plants in the salt water tank, we left several sprouts in vermiculite without moving to the salt water tank, and cultivated them with 2\% NaCl water (NaHCO$_3$ was not included). Although growth rate was different between each plant, several sprouts grew higher than the plants in the salt water tank. While ice plant hates salt even though it is tolerant, salicornia seems to use salt to increase its growth. Although the growth speed is slow, salicornia may best match with Enceladus' water of three species.

\section{Summary}
So far, Enceladus is one of the most probable celestial bodies that have the subsurface ocean (e.g., Porco et al., 2006; Postberg et al., 2009, 2011). From the analyses of observed plume, Enceladus' ocean seems to contain salts such as NaCl and carbonate (Postberg et al., 2009, 2011). In order to test whether plants can grow with Enceladus' water, by hydroponic, we tried to cultivate three salt-tolerant plants (ice plant, swiss chard, salicornia) with 0.33\% NaCl and 0.4\% NaHCO$_3$ using table salt and baking soda. If sprouts germinated with pure water were moved to the salt water, all three plants could grow. However, compared with the plants cultivated with pure water, growth can be suppressed in the case of ice plant. Swiss chard also grew in the salt water. However, the base of the stem became brown in the salt water tank. Because it fell several times, salt in water might damage it. Salicornia does not seem to be suppressed to grow by salt water. However, the growth speed is slow. For 1.5 months, sprouts of salicornia grew only by a few centimeters. These are the preliminary results of the first trial. Obviously, in order to test plant growth with Enceladus' water, more works are required. Water made in this work has pH$\sim$8, which is smaller than the value suggested for Enceladus' ocean (pH=8.5-10.5) (Hsu et al., 2015). We are planning to use ammonia or Na$_2$CO$_3$ instead of NaHCO$_3$ to make high pH in future studies. Tests with different salinity also should be considered.

In planetary science, for the analysis of morphology of celestial bodies, terrestrial analogue sites are usually referenced. For example, for analysis of subsurface ocean in icy bodies such as Europa and Enceladus, Lake Vostok in antarctica can be an analogue site. On the Earth, cultivation of salt-tolerant plants can be one solution to the agriculture on salinized ground (e.g., Rozema and Flowers, 2008). As "opposite" analogues, researches on growing plants outside the Earth may be referenced for analysis of terrestrial cultivation. If so, Enceladus can be one important analogue site for terrestrial botany and agriculture. 

\section*{Acknowledgements}
Y. Sekine gave comments about salinity of Enceladus' water.

\newpage

\end{document}